\documentclass[traditabstract]{aa} 
\usepackage{graphicx}
\usepackage{txfonts}

\def\0{\phantom0}

\def\obj{SDSS~J0806+2006}

\def\msol{M$_{\odot}$}

\begin{document}
   \title{A sharp look at the gravitationally lensed quasar \obj\ with
   laser  guide  star  adaptive optics   at the  VLT{\thanks{Based  on
   observations  collected   at the   European Southern   Observatory,
   Paranal, Chile, ESO Program 079.A-0588(A).}}}


\titlerunning{Laser guide star adaptive optics imaging of \obj}

   \author{D.~Sluse\inst{1} \and F.~Courbin\inst{1} \and A.~Eigenbrod \inst{1} 
\and G.~Meylan\inst{1}}

   \institute{Laboratoire d'astrophysique, Ecole Polytechnique
     F\'ed\'erale de Lausanne (EPFL), Observatoire de Sauverny,
     CH-1290 Versoix, Switzerland} 

   \date{Received 17 September 2008; Accepted 6 November 2008}

 
   \abstract{We present the first VLT near-IR observations of a
     gravitationally lensed quasar, using adaptive optics and laser
     guide star.  These observations can be considered as a test bench
     for future systematic observations of lensed quasars with
     adaptive optics, even when bright natural guide stars are not
     available in the nearby field.  With only 14 minutes of observing
     time, we derived very accurate astrometry of the quasar images
     and of the lensing galaxy, with 0.05\arcsec\ spatial resolution,
     comparable to the Hubble Space Telescope (HST). In combination
     with deep VLT optical spectra of the quasar images, we use our
     adaptive optics images to constrain simple models for the mass
     distribution of the lensing galaxy. The latter is almost circular
     and does not need any strong external shear to fit the data. The
     time delay predicted for \obj, assuming a singular isothermal
     ellipsoid model and the concordance cosmology, is $\Delta t
     \simeq 50$ days.  Our optical spectra indicate a flux ratio
     between the quasar images of A/B=1.3 in the continuum and A/B=2.2
     in both the \ion{Mg}{ii} and in the \ion{C}{iii]} broad emission
     lines.  This suggests that microlensing affects the continuum
     emission. However, the constant ratio between the two emission
     lines indicates that the broad emission line region is not
     microlensed. Finally, we see no evidence of reddening by dust in
     the lensing galaxy.}


   \maketitle
%

\section{Introduction}

Strong gravitational lensing of distant quasars is a useful tool both
for cosmology and for studying luminous and dark matter in massive
galaxies.  For a fixed cosmology, the total projected distribution of
matter in the galaxy(ies) responsible for the lensing effect can be
reconstructed from (i) the configuration of the lensed quasar images
on the plane of the sky, (ii) the observed magnification ratio between
the quasar images, (iii) measurement of the time delay, and (iv)
detailed structures in the lensed host galaxy of the quasar forming
arcs and counter-arcs (e.g. Refsdal~\cite{REFS64}, Claeskens \&
Surdej~\cite{CLA02}, Kochanek et al.~\cite{KOC06}).  Finally, the
magnification ratios between the quasar images can significantly
deviate from the predictions by smooth mass models if the lensing
galaxy contains small-scale substructures with $M\sim 10^{8-9}\,$\msol\,
(e.g., Mao \& Schneider~\cite{MAO98}, Metcalf~\cite{MET05}).  Strong
gravitational lensing of quasars is therefore a sensitive indirect way
of detecting small satellites of lensing galaxies.

High-resolution imaging of lensed quasars is crucial for constraining
mass models for the lensing galaxy, determining its luminous mass
distribution, and identifying the optical counterpart of any mass
substructure (Schechter \& Moore ~\cite{SCH93}, Koopmans \&
Treu~\cite{KOO02}, McKean et al.~\cite{MCK07}). The Hubble Space
Telescope (HST) has provided the community with a large sample of
exquisite high-resolution images of gravitational lenses with the
CASTLES survey (CfA-Arizona Space Telescope LEns Survey; Mu{\~n}oz et
al.~\cite{MUN98}) and played a crucial role in the study of lensed
quasars (Claeskens et al.~\cite{CLA07} for a review). Ground-based
adaptive optics (AO) imaging now allows us to obtain high spatial
resolution data from the ground.  Although both approaches are
complementary in depth and field of view, the AO imaging data
collected so far of lensed quasars are still very scarce.  With
Gemini-North, Courbin et al.~(\cite{COU02}) obtained the first AO
images of a gravitational lens, for the Einstein ring
PKS~1830-21. These observations were followed up at the VLT in AO in
order to confirm a double lensing galaxy (Meylan et al.~\cite{MEY05}).
More recently, Auger et al.~(\cite{AUG08}) used AO to observe the
doubly imaged quasar SBS~1520+53.  These two targets are easy for AO,
as they are located a few arcseconds away from a bright natural guide
star.  Systematic AO imaging of all other lensed quasars requires
laser guide star (LGS).  McKean et al.~(\cite{MCK07}) obtained the
first AO/LGS images at the Keck of a lensed quasar to search for
substructures in the lensing galaxy.

In the present paper, we describe the first AO/LGS images of a
gravitational lens at the VLT: the doubly imaged quasar \obj, at
$z_s=1.540$. This quasar has been found to be gravitationally lensed
from a systematic search for strong lens systems in the Sloan Digital
Sky Survey (Inada et al.~\cite{INA06}).  The angular separation
between the two quasar images is $\Delta \theta \sim 1.5\arcsec$ and
the early-type lensing galaxy has been found to be at $z_l=0.573$ from
deep VLT optical spectroscopy (Eigenbrod et al.~\cite{EIG07}).  Our
new VLT AO/LGS images, further improved with image deconvolution,
allow us to derive milli-arcsecond astrometry of the quasar images, as
well as improved models for the lensing galaxy. We see these
observations as a test bench for systematic VLT AO imaging of
gravitationally lensed quasars.


\begin{figure*}[ht!]
\begin{center}
\includegraphics[width=14.0cm]{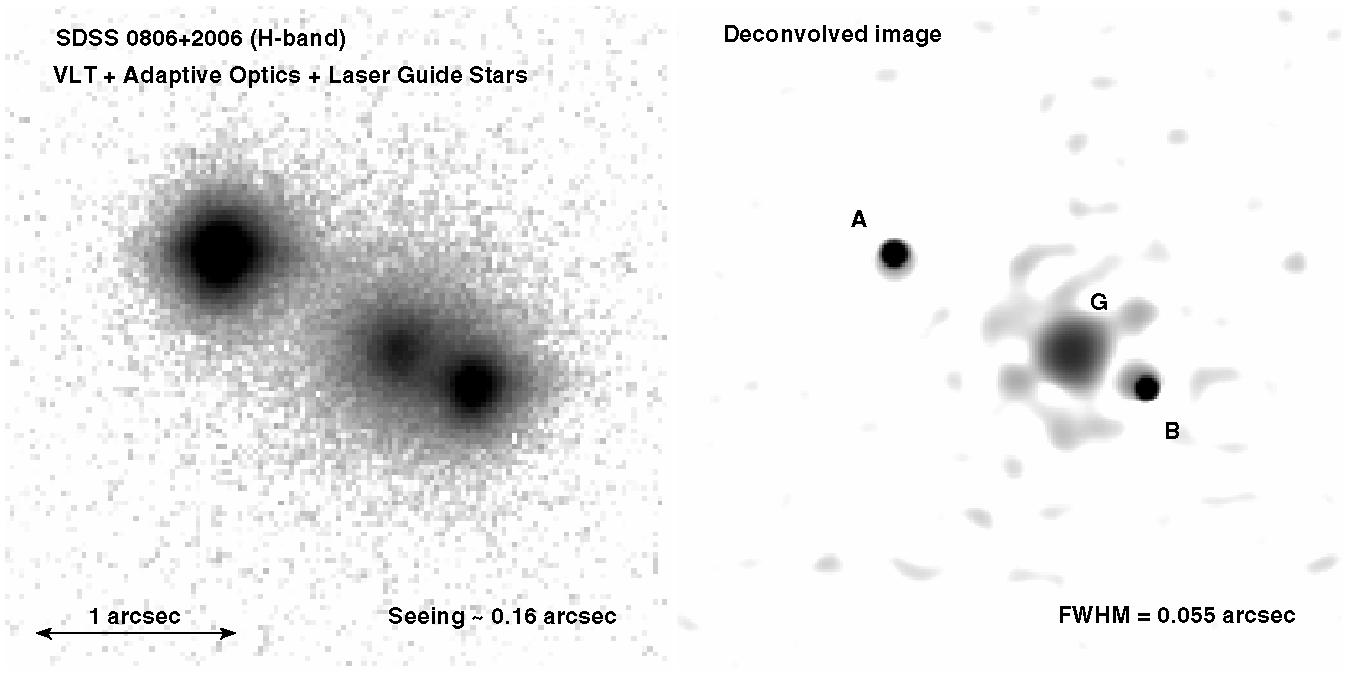}
\caption{{\it Left:} VLT image of \obj\, obtained with the adaptive
  optics + laser guide star facility.  The PSF FWHM in this 14-minute
  $H$-band image is about the same as on a HST/NICMOS image in the
  same band.  {\it Right:} deconvolved image, as obtained with the MCS
  algorithm (see text).  The lower cut level in this deconvolved image
  is set to 1$\sigma_{\rm sky}$ in order to show the noise
  level. North is to the top, east to the left.}
\label{fig:ima}
\end{center}
\end{figure*}

\section{Observations and reduction}

\subsection{VLT adaptive optics observations}

Our AO/LGS $H$-band images of \obj\ were obtained in service mode at
the ESO Paranal Observatory with the near-infrared camera CONICA (the
COude Near-Infrared CAmera), which is mounted on the AO system NAOS
(Nasmyth Adaptive Optics System), installed at the Nasmyth B focus of
the VLT-UT4. This instrument uses an LGS Facility, in operation at
Paranal Observatory since April 2007.  The LGS is an artificial source
replacing the natural guide star as a reference object for AO image
correction.  A natural guide star is still required to correct for the
tip-tilt motions, which are not sensed by the LGS, but the natural
guide star can be much fainter ($12 < V < 17$) than without the
LGS. In addition, the natural guide star can be as far as 55$\arcsec$
away from the science target.  The S27 CONICA camera was used, with a
pixel size of 27.053$\pm$0.019 mas, resulting in a field of view of
28$\times$28\arcsec (Amico et al.~\cite{AMI08}).  A set of 7 exposures
of 120\,s each were obtained in the $H-$band on 2008 January 7, at a
mean airmass of $sec(z) \sim$ 1.4, under seeing conditions in the
range 0.65\arcsec-1.0\arcsec.  We used the star U0975\_07317639
(V=14.2), located 44.6\arcsec\ from \obj , as a reference tip-tilt
star.  These relatively poor observing conditions led to a Strehl
ratio on-target of only 6\%, with a final full width half at maximum
spatial resolution of 0.16$\arcsec$ on the coadded image.  The point
spread function (PSF) varied from frame to frame, with an FWHM varying
from 0.13\arcsec\ to 0.19\arcsec\ and an ellipticity varying between
0.0 and 0.2.

Standard near-IR reduction procedures were applied to subtract the
dark and to flat-field the images using a normalized twilight flat
field. The sky subtraction and the co-addition of the reduced dithered
frames were performed using the {\texttt{xdimsum}} IRAF{\footnote{IRAF
    is distributed by the National Optical Astronomy Observatories,
    which are operated by the Association of Universities for Research
    in Astronomy, Inc., under cooperative agreement with the National
    Science Foundation.}} package. The sky frames were computed using
the 3 images taken nearest in time to the frame considered.  The
standard star S301-D, observed on the same night as \obj, was reduced
in a similar way to the science data. From aperture photometry, we
derived a zeropoint $H=23.87 \pm 0.10$ mag/s.  In this calculation we
adopted a Paranal extinction coefficient of 0.06 mag per unit airmass
in the $H-$band. The large uncertainty on the zeropoint reflects the
inaccuracy of absolute photometry with AO systems due to seeing
variations, PSF halo variations and angular anisoplanatism
(e.g. Esslinger \& Edmunds ~\cite{ESS98}). The error on the relative
photometry between images A and B of \obj\, should be less than
0.05 mag (Esslinger \& Edmunds~\cite{ESS98}) due to the small PSF
variation over the 2\arcsec$\times$2\arcsec field covered by the lens
system.

\subsection{VLT spectroscopy}

A set of 4 low-resolution spectra of \obj\ is available, obtained with
the FORS1 instrument mounted on the ESO-VLT at the Paranal Observatory
(Chile).  Using these data, taken on 2006 April 22, Eigenbrod et al.
(\cite{EIG07}) measured the redshift of the lensing galaxy. In the
following, we analyze the spectra again in order to infer the broad
emission line flux ratio of the quasar images and to estimate the dust
content of the lensing galaxy.  The details of the observations and
reduction of the spectra can be found in Eigenbrod et
al.~(\cite{EIG06a, EIG06b, EIG07}).

All of the observations were carried out using multi-object spectroscopy
(MOS). The object slit encompasses images A-B and the lensing
galaxy. The other slits were placed on 4 stars, that were used as flux
calibrators and as reference PSFs in order to deblend the quasar
images. The wavelength range 4450 $< \lambda < $ 8650 \AA\, was covered
with a resolving power R = $\lambda/\Delta \lambda=200$ at the central
wavelength.

The final combined 2D spectrum of \obj\ was spatially deconvolved using
the spectral version of the MCS deconvolution algorithm (Courbin et
al. \cite{COU00}; Sect.~\ref{sec:spectro}), leading to individual 1D
quasar spectra, free of any mutual contamination. These spectra were
eventually corrected from the spectrograph response curve and from
differential atmospheric extinction using ta\-bu\-la\-ted Paranal
extinction curves.

\section{Analysis}

\subsection{Imaging: deconvolution and profile fitting}
\label{sec:imaging}

\begin{table}[b!]
  \centering 
  \caption{Results of the fit of the lensing galaxy with \texttt{GALFIT} 
for 2 different light profiles.   }
  \vspace{0.2cm}
  \begin{tabular}{c||ccccc}
    \hline
    \hline
Model & $\chi^2$ & d.o.f. & PA & $q=b/a$ & $R_e$ (\arcsec) \\ 
    \hline
GdV & 1.06 & 16372 & 78.2 $\pm$ 6.8 & 0.88 $\pm$ 0.02 & 8.9$\pm$0.3 \\
Exp & 1.03 & 16372 & 83.9 $\pm$ 6.8 & 0.91 $\pm$ 0.01 & 7.0$\pm$0.1 \\
    \hline
  \end{tabular}
  \label{tab:galax}
\end{table}

\begin{table}[b!]
  \centering 
  \caption{Relative astrometry and  $H$-band photometry for the lensed images A and B 
    and for the lensing galaxy G. }
  \vspace{0.2cm}
  \begin{tabular}{c||ccc}
    \hline
    \hline
    ID & $\Delta \alpha$ (\arcsec) & $\Delta \delta$ (\arcsec) & $H$-Magnitude \\ 
    \hline
    A  & $+0.000 \pm 0.003 $ & $ +0.000 \pm 0.003$ & 17.19 $\pm$ 0.14 \\
    B  & $-1.316 \pm 0.003 $ & $ -0.701 \pm 0.003$ & 18.00 $\pm$ 0.14 \\
    G  & $-0.918 \pm 0.009 $ & $ -0.514 \pm 0.009$ & 17.56 $\pm$ 0.18 \\
    \hline
  \end{tabular}
  \label{tab:astrom}
\end{table}

We analyzed our AO/LGS images in two ways. First, we applied the MCS
image deconvolution algorithm (Magain et al. \cite{magain}).  As no
star was available in the immediate vicinity of \obj, we built the PSF
on the relatively isolated quasar image~A (Fig.~1). The resolution in
the final deconvolved image was chosen to be 0.055\arcsec\ FWHM. The
frame was decomposed into an analytical channel containing the point
sources and a numerical channel containing an image of the lensing
galaxy. The result of the deconvolution technique is shown in the
right panel of Fig.~1. We measured the position of the lensing galaxy
using Gaussian fitting on the numerical channel of the
deconvolution. The astrometry and photometry of the quasar images were
returned as analytical parameters of the point-source channel of the
deconvolution.

Second, we also analyzed the images by modeling the quasar images and
the galaxy light distributions using \texttt{GALFIT} (v2.0.3c; Peng et
al. \cite{PEN02}). For this purpose we used quasar image A as our PSF
model and simultaneously fit the 2 quasar images and an analytical
galaxy profile, convolved with the PSF. The parameters of the lensing
galaxy are displayed in Table~\ref{tab:galax} for de Vaucouleurs (GdV)
and exponential-disk (Exp) light profiles. Both models lead to equally
good fits to the data, given our signal-to-noise, with a slightly
better formal fit for the exponential profile. However, as the
spectrum of the lensing galaxy is typical of an early type galaxy
(Eigenbrod et al. \cite{EIG07}), we use in the following the
morphological properties derived from the de Vaucouleurs model. In
Table~\ref{tab:galax}, we give the reduced $\chi^2$, the number of
degrees of freedom (d.o.f.), the galaxy PA (east of north), the axis
ratio $q$, and the effective radius $R_e$.  The error bars associated
to the galaxy parameters are formal error bars of the fit and do not
include systematic errors associated with the choice of surface
brightness profile.  These systematic errors are likely to be dominant for
the effective radius, $R_e$, and axis ratio, $q$, as well as for the
magnitude of the galaxy

The deconvolution procedure and the \texttt{GALFIT} minimization lead
to photometric measurements compatible within 0.05 mag and astrometric
measurements compatible within 0.005\arcsec. We summarize the
astrometry and photometry of \obj\, in Table~\ref{tab:astrom}. The
photometry is in the Vega system. The astrometry of the point sources
is given by the deconvolution process, while the position of the
lensing galaxy is given by \texttt{GALFIT}. Measuring the position of
the numerical lensing galaxy of the MCS deconvolution gives consistent
results. The error bars quoted in Table~\ref{tab:astrom} are $1
\sigma$ error bars including the estimated systematic errors.

We find that the separation between the quasar images is 0.1\arcsec\,
more than the one presented in Inada et al.~(\cite{INA06}),
incompatible with their quoted error bars. An overestimate by about
6\% of the CONICA pixel size would allow both data measurements to be
reconciled. However, the CONICA pixel size has been recently confirmed
by new measurements from Sicardy et al. (private communication) and by
our measurements of the relative astrometry of 4 stars in the field of
\obj.

\begin{table*}[t!]
  \centering 
  \caption{Results of the lens models, where angles are measured in degrees 
east of north. }
  \vspace{0.2cm}
  \begin{tabular}{cccccccccc}
    \hline
    \hline
    Model & $R_E$(\arcsec) & $\gamma$ or $e$ & $\theta_{\gamma}$ or $\theta_e$ & $\Delta t $($h^{-1}$ days) & $\kappa_A$ & $\gamma_A$ &  $\kappa_B$ & $\gamma_B$    \\
    \hline
SIS+$\gamma$ & 0.74 & 0.017 & 56.8 & 36.3 & 0.351 & 0.342 & 0.841 & 0.833 \\
SIE & 0.76 & 0.042 & 76.6 & 36.2 & 0.356 & 0.356 & 0.850 & 0.850 \\ 

    \hline
  \end{tabular}
  \label{tab:mod}
\end{table*} 

\subsection{Spectroscopy}
\label{sec:spectro}

\begin{figure}[t!]
\begin{center}
\includegraphics[width=9.0cm]{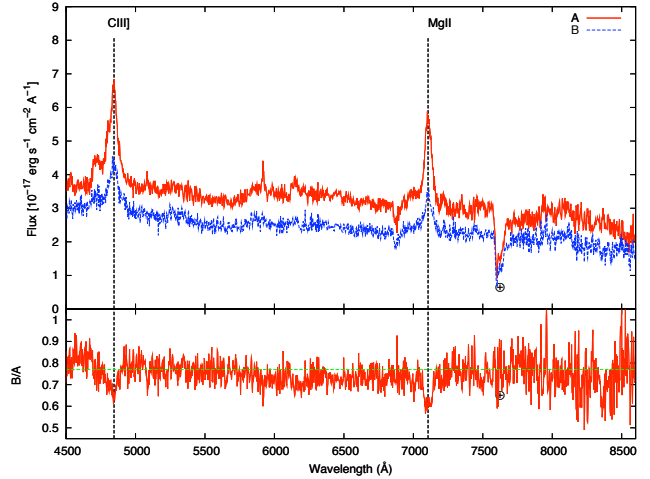}
\caption{{\it Top:} our deep VLT optical spectra of components A (dotted blue) 
and B (dashed red) of \obj. {\it Bottom:} spectral ratio between B and
A.  The straight line  at B/A =0.77  (A/B=1.3) shows the average ratio
estimated for the continuum emission. }
\label{fig:spectro}
\end{center}
\end{figure}

The original deconvolution of the 2D spectrum of \obj\ presented by
Eigenbrod et al. (\cite{EIG07}) is aimed at measuring the redshift of
the lensing galaxy.  We optimize here the deconvolution in order
  to improve the signal-to-noise ratio in the spectra of the quasar
  images and minimize the contamination of the point-like sources by
  the extended lensing galaxy. 

The spectra of the quasar images are shown in
Fig.~\ref{fig:spectro}.  Although the conditions were photometric
during the observations, the spectra are not corrected from the flux
loss due to the finite slit width. Still, the relative fluxes are correct
since the slit clipping is the same for the two quasar images and
since FORS1 uses an atmospheric refraction corrector.

The spectral ratio between the quasar images provides a good
diagnostic of the presence of differential microlensing and
differential extinction (e.g. Sluse et al. \cite{SLU07}). The imprint
of quasar broad emission lines (BEL) is clearly visible in $B/A$
(Fig.~\ref{fig:spectro}), which is otherwise flat over the observed
wavelength range.  This is a clear signature of differential
microlensing between the continuum emission region and the broad
emission line region. Indeed, the Einstein radius of a microlens
located in the lensing galaxy is about $R_{\mu E} =$ 0.0012
$h^{-1/2}\sqrt{M/M_{\odot}}$ pc (projected in the source plane).
Consequently, the continuum emission region, with a size $<10^{-4}$\,pc $<
R_{\mu E}$, is more likely microlensed than the much larger broad
emission line region. The flux ratio in the continuum is found to be
$A/B(cont)=1.3$, while the ratio in the \ion{Mg}{ii}\,and
\ion{C}{iii]}\, is found to be $A/B(BEL)=2.2$.  Interestingly, the BEL
measurement also agrees with the flux ratio measured in the
$H-$band, i.e., $A/B(H)=2.1$.  This is consistent with the need for
the $H-$band continuum emission to arise from a larger region than
the optical continuum emission. Like Inada et al.~(\cite{INA06}), we
do not detect any significant chromatic change of $A/B$ in our
spectra.  This indicates that differential extinction between the
quasar lensed images is small.  Although microlensing could induce
chromatic variations in the quasar flux ratios (Wambsganss \&
Paczy\`nski~\cite{WAM91}), the absence of such changes in microlensed
quasars is not uncommon.

Finally, there is an apparent change in the flux ratio $A/B$ between
April 2005 and April 2006.  Inada et al.~(\cite{INA06}) report
$A/B(cont)\sim 1.45$ in the continuum spectral ratio and in the
optical images, while we measure $A/B(cont)=1.3$.  This is compatible
with an increase of 10\% in the micro-magnification between April 2005
and April 2006. Systematic errors in the spectral extraction and in
the image flux measurements, along with the combined effects of
intrinsic variability and a time delay might also explain the observed
difference.

\section{Lens Modeling}

Due to the mismatch between the astrometry exposed in
Table~\ref{tab:astrom} and the one in the discovery paper (Inada et
al.~\cite{INA06}), we present new mass models of the lensing galaxy.
Two standard mass models are used: a singular isothermal sphere with
an external shear (SIS+$\gamma$) model and a singular isothermal
ellipsoid (SIE) model. Both models have eight parameters: the Einstein
radius $R_E$, the shear $\gamma$ or ellipticity $e$ and their
respective position angles ($\theta_{\gamma}$ or $\theta_e$), the
position of the lens galaxy, and the position and flux of the source
quasar.  We have 8 nominal constraints (the positions of A, B, and G,
the flux of A and B). The PA of the galaxy light distribution can also
be used to constrain the SIE mass model.  Indeed, statistical studies
find correlations between the position angles but not between the axis
ratios of the visible and total mass distributions (Keeton et al.
\cite{KEE97}, Ferreras et al.  \cite{FER08}).  In conclusion, the
SIS+$\gamma$ model has 0 d.o.f., while the SIE model has 1 d.o.f. We
used the quasar and galaxy positions reported in
Table~\ref{tab:astrom} and the flux ratio $B/A=$ 0.455 $\pm$ 0.050
measured in the broad emission lines.  We chose this flux ratio
because it is less affected by microlensing than the continuum flux
ratio.

We fit the data using \texttt{LENSMODEL} (v1.99g) (Keeton
\cite{KEE01}). The results are summarized in Table~\ref{tab:mod}. In
addition to the model parameters, we give the local convergence
$\kappa$ and shear $\gamma$ at the location of the lensed images as
derived from the magnification tensor.  The SIS+$\gamma$ model fits
the data perfectly, as expected (d.o.f.= 0).  The shear is found to be
very small ($\gamma=$0.017). The reduced $\chi^2$ for the SIE model is
$\chi^2=0.81$.  This model shows that there is excellent agreement
between the PA of the light and of the mass distributions.  These
results differ significantly from those presented in Inada et
al.~(\cite{INA06}).  The origin of this discrepancy is the
significantly different relative astrometry of \obj\ obtained with our
AO imaging.

\section{Conclusions}

We obtained the first VLT images of a gravitationally lensed
quasar using AO and LGS. Using image deconvolution we derived accurate
astrometry of the quasar images and of the lensing galaxy. This
astrometry is significantly different from the one in Inada et
al.~(\cite{INA06}) and is not due to a systematic error in the pixel
size of the instrument used.

We  find that a  simple isothermal mass  model can account for the new
astrometry, with no need of strong external shear.  The predicted time
delay for this model, $\Delta t \simeq 36.3\,h^{-1}$  days, is well-suited to
an   accurate   measurement  using   photometric  light    curves (see
simulations by Eigenbrod et al. \cite{EIG05}).

Our VLT optical spectra show that \obj\, is affected neither by
chromatic microlensing nor by dust extinction. The continuum flux
ratio based on these spectra agrees within 10\% with the one
derived by Inada et al. one year before, suggesting low-amplitude
microlensing-induced variations.  \obj\, is therefore a clean lens
system to use either for determining of $H_0$ or for the study
of the mass distribution in the lensing galaxy, based on a time-delay
measurement.

The accuracy of the relative positions of the lensed images derived
from our observations of \obj\ is similar to the typical accuracy
derived from the HST data of other lensed quasars (e.g. Impey et
al.~\cite{IMP98}, Claeskens et al. ~\cite{CLA06}). This demonstrates
that ground-based AO imaging with the help of LGS can be fully
competitive with HST observations in the near-IR. About 90\% of the
lensed quasars observable with the VLT are either bright enough to be
used as a tip-tilt reference or have potential tip-tilt stars with V
$<$ 17 within a 55\arcsec radius.  This is promising for future systematic
observations of gravitational lenses with adaptative optics and laser
guide star.

\begin{acknowledgements}

  We thank C. Lidman, N. Ageorges, P. Amico and L. Tacconi-Garman for
  extensive discussions concerning CONICA pixel size and use of the
  LGS facility. We also thank B. Sicardy for providing ESO with the
  new CONICA pixel size prior to publication. We are indebted to the
  anonymous referee for the valuable comments on the first version of
  this letter. This work is partly supported by the Swiss National
  Science Foundation (SNSF).
\end{acknowledgements}

\end{document}